\documentclass{article}

\usepackage{pdfpages}
\usepackage{fullpage}
\usepackage{textcomp}
\usepackage{algorithm}
\usepackage{algorithmicx}
\usepackage{algpseudocode}
\usepackage{xcolor}
\usepackage{multirow}
\usepackage{amsmath}
\usepackage{amsfonts}
\usepackage{graphicx}
\usepackage{amssymb}
\usepackage{amsthm}

\newtheorem{lemma}{Lemma}
\newtheorem{theorem}[lemma]{Theorem}

\newtheorem{problem}[lemma]{Problem}

\newcommand{\IR}{\mathbb{R}}
\newcommand{\tO}{\tilde{O}}

\title{Range closest-pair search in higher dimensions%
  \footnote{Work by the first author has been partially supported by NSF Grant CCF-1814026. Work by the third author has been partially supported by a Doctoral Dissertation Fellowship from the Graduate School of the University of Minnesota.}
}

\author{Timothy M. Chan \footnote{University of Illinois at Urbana-Champaign, Urbana, IL, USA} \\ \texttt{tmc@illinois.edu}
\and Saladi Rahul \footnotemark[2] \\ \texttt{saladi.rahul@gmail.com}
\and Jie Xue \footnote{University of Minnesota - Twin Cities, Minneapolis, MN, USA} \\ \texttt{xuexx193@umn.edu}}

\date{}

\begin{document}

\maketitle

\begin{abstract}
Range closest-pair (RCP) search is a range-search variant of the classical closest-pair problem, which aims to store a given set $S$ of points into some space-efficient data structure such that when a query range $Q$ is specified, the closest pair in $S \cap Q$ can be reported quickly.
RCP search has received attention over years, but the primary focus was only on $\mathbb{R}^2$.
In this paper, we study RCP search in higher dimensions.
We give the first nontrivial RCP data structures for orthogonal, simplex, halfspace, and ball queries in $\mathbb{R}^d$ for any constant $d$.
Furthermore, we prove a conditional lower bound for orthogonal RCP search 
for $d \geq 3$.
\end{abstract}

\section{Introduction}
The closest-pair problem is one of the most fundamental problems in computational geometry and finds numerous applications in various areas, such as collision detection, traffic control, etc.
In many scenarios, instead of finding the global closest-pair, people want to know the closest pair contained in some specified ranges.
This results in the notion of \textit{range closest-pair} (RCP) search.
RCP search is a range-search variant of the classical closest-pair problem, which aims to store a given set $S$ of points into some space-efficient data structure such that when a query range $Q$ is specified, the closest pair in $S \cap Q$ can be reported quickly.
RCP search has received considerable attention over the years \cite{abam2013power,bae2018closest,gupta2014data,gupta2006range,shan2003spatial,sharathkumar2007range,xue2018approximate,xue2019colored,xue2018new,xue2018searching}.

Unlike most traditional range-search problems, RCP search is \textit{non-decomposable}.
That is, if we partition the dataset $S$ into $S_1$ and $S_2$, given a query range $Q$, the closest pair in $S \cap Q$ cannot be obtained efficiently from the closest pairs in $S_1 \cap Q$ and $S_2 \cap Q$.
Due to the non-decomposability, many traditional range-search techniques are inapplicable to RCP search, which makes the problem quite challenging.
As such, despite of much effort made on this topic, most known results are restricted to the plane case, i.e., RCP search in $\mathbb{R}^2$.
Beyond $\mathbb{R}^2$, only very specific query types have been studied, such as 2-sided box queries.

In this paper, we investigate RCP search in higher dimensions.
We consider four widely-studied query types: orthogonal queries, simplex queries, halfspace queries, and ball queries.
We are interested in designing efficient RCP data structures (in terms of space cost, query time, and preprocessing time) for these kinds of query ranges, and proving conditional lower bounds for these problems.


\medskip\noindent
\textbf{Related work.} \label{sec-work}
The closest-pair problem and range search are both well-studied problems in computational geometry; see \cite{agarwal1999geometric,smid1995closest} for surveys of these two topics.

RCP search was for the first time introduced by Shan et al.~\cite{shan2003spatial} and subsequently studied in \cite{abam2013power,bae2018closest,gupta2014data,gupta2006range,sharathkumar2007range,xue2018approximate,xue2019colored,xue2018new,xue2018searching}.
In $\mathbb{R}^2$, the query types studied include quadrants, strips, rectangles, and halfplanes.
RCP search with these query ranges can be solved using near-linear space with poly-logarithmic query time.
The best known data structures were given by Xue et al.~\cite{xue2018new}, and we summarize the bounds in Table~\ref{tab-2Dresult}.
For \textit{fat} rectangles queries (i.e., rectangles of constant aspect ratio), Bae and Smid \cite{bae2018closest} showed an improved RCP data structure using $O(n \log n)$ space and $O(\log n)$ query time.
In a recent work \cite{xue2019colored}, Xue considered a colored version of RCP search in which the goal is to report the bichromatic closest pair contained in a query range, and proposed efficient data structures for orthogonal colored approximate RCP search (mainly in $\mathbb{R}^2$).

\begin{table}[h]
    \centering
    \begin{tabular}{c|c|c|c}
        \hline
        Query type & Space cost & Query time & Preprocessing time  \\
        \hline
        Quadrant & $O(n)$ & $O(\log n)$ & $O(n \log^2 n)$ \\
        \hline
        Strip & $O(n \log n)$ & $O(\log n)$ & $O(n \log^2 n)$ \\
        \hline
        Rectangle & $O(n \log^2 n)$ & $O(\log^2 n)$ & $O(n \log^7 n)$ \\
        \hline
        Halfplane & $O(n)$ & $O(\log n)$ & $O(n \log^2 n)$ \\
        \hline
    \end{tabular}
    \caption{Best known results in $\mathbb{R}^2$}
    \label{tab-2Dresult}
\end{table}

Beyond $\mathbb{R}^2$, the problem is quite open.
To our best knowledge, the only known results are the orthogonal RCP data structure given by Gupta et al.~\cite{gupta2014data} which only has guaranteed average-case performance and the approximate colored RCP data structures given by Xue \cite{xue2019colored} which can only handle restricted query types (dominance query in $\mathbb{R}^3$ and 2-sided box query in $\mathbb{R}^d$).

A key ingredient in existing solutions for RCP search in $\mathbb{R}^2$ is the \textit{candidate-pair} method.
Roughly speaking, this method tries to show that among the $\Omega(n^2)$ point pairs, only a few (called \textit{candidate pairs}) can be the answer of some query.
If this can be shown, then it suffices to store the candidate pairs and search the answer among them.
Unfortunately, it is quite difficult to generalize this method to higher dimensions, as the previous approaches for proving the number of candidate pairs heavily rely on the fact that the data points are given in the plane.
This might be the main reason why RCP search can be efficiently solved in $\mathbb{R}^2$, while remaining open in higher dimensions.

\medskip\noindent
\textbf{Our contributions.} \label{sec-contribution}
In this paper, we give the first non-trivial RCP data structures for orthogonal, simplex, halfspace, and ball queries in $\mathbb{R}^d$, for any constant $d$.
The performances of our new data structures are summarized in Table~\ref{tab-HDresult}, where the notation $\tilde{O}(\cdot)$ hides $\log n$ factors.
All these data structures have near-linear space cost, sub-linear query time, and sub-quadratic preprocessing time.
For example, we obtain $\tO(n^{7/8})$ query time for two-dimensional triangular ranges, and $\tO(n^{2/3})$ query time for three-dimensional halfspaces and two-dimensional balls (i.e., disks).\footnote{Gupta et al.~\cite{gupta2014data} obtained $\tO(\sqrt{n})$ query time for two-dimensional disks, but only for \emph{uniformly distributed} point sets; the general problem was left open in their paper.}

Furthermore, we complement these results by establishing a conditional lower bound, implying that our $\tO(\sqrt{n})$ query time bound for orthogonal RCP search in $\mathbb{R}^d$ for any $d \geq 3$ is likely the best possible (and in particular explaining why polylogarithmic solution seems not possible beyond two dimensions).
Specifically, we show that orthogonal RCP search in $\mathbb{R}^3$ is at least as hard as the \emph{set intersection query problem}, which is conjectured to require $\tilde{\Omega}(\sqrt{n})$ query time for linear-space structures.

\begin{table}[h]
    \centering
    \begin{tabular}{c|c|c|c|c}
        \hline
        Query type & Source & Space cost & Query time & Preprocessing time  \\
        \hline
        Orthogonal & Theorem~\ref{thm:orthogonal-ub} & $\tilde{O}(n)$ & $\tilde{O}(\sqrt{n})$ & $\tilde{O}(n\sqrt{n})$ \\
        \hline
        Simplex & Theorem~\ref{thm-SRCP} & $\tilde{O}(n)$ & $\tilde{O}(n^{1-1/(2d^2)})$ & $\tilde{O}(n^{(3d^2+1)/(2d^2+1)})$ \\
        \hline
        Halfspace & Theorem~\ref{thm-HRCP} & $\tilde{O}(n)$ & $\tilde{O}(n^{1-1/(d \lfloor d/2 \rfloor)})$ & $\tilde{O}(n^{2-1/(2d^2)})$ \\
        \hline
        Ball & Theorem~\ref{thm-BRCP} & $\tilde{O}(n)$ & $\tilde{O}(n^{1-1/((d+1) \lceil d/2 \rceil)})$ & $\tilde{O}(n^{2-1/(2(d+1)^2)})$ \\
        \hline
    \end{tabular}
    \caption{Performances of our new RCP data structures in $\mathbb{R}^d$}
    \label{tab-HDresult}
\end{table}

        
\noindent
\textbf{Overview of our techniques.}
Our approach for designing these new data structures is quite different from those in the previous work.
We avoid using the aforementioned candidate-pair method.
Instead, our RCP data structures solve the problems as follows (roughly).
For a given query range $Q$, the data structure first partitions the points in $S \cap Q$ into two subsets, say $K$ and $L$.
The size of $L$ is guaranteed to be small, while $K$ may have a large size.
Then the data structure computes the closest pair $\phi$ in $K$ using some pre-stored information and computes the closest pair $\phi'$ in $L$ using the standard closest-pair algorithm (which can be done efficiently as $L$ is small).
If the two points of the closest pair $\phi^*$ in $S \cap Q$ are both in $K$ or both in $L$, we are done.
The only remaining case is that one point of $\phi^*$ is in $K$ while the other point is in $L$.
The data structure handles this case by finding the nearest neighbor of $a$ in $Q$ for every $a \in L$ via reporting all the points in $Q$ that are ``near'' $a$.
Using a packing argument, we can show that one only needs to report a constant number of points for each $a \in L$, and hence this procedure can be completed efficiently (since $L$ is small).

To implement this strategy, we incorporate a number of existing geometric data structuring techniques. For orthogonal RCP, we use \emph{range trees} and adapt an idea from Gupta et al.~\cite{gupta2014data} of classifying nodes as ``heavy'' and ``light'' (originally for solving a different problem, two-dimensional orthogonal \emph{range diameter}, in near-linear space and $\tO(\sqrt{n})$ query time).  For simplex RCP, we use \emph{simplicial partitions} instead of range trees.  For halfspace RCP, we switch to dual space and use \emph{cuttings}, similar to an idea from Chan et al.~\cite{cdlmw14} (for solving a different problem, halfspace \emph{range mode}, in near-linear space and $\tO(n^{1-1/d^2})$ time).  Overall, the combination of existing and new ideas is nontrivial (and interesting, in our opinion).
Our conditional lower bound proof for three-dimensional orthogonal RCP is similar to some previous work (for example, Davoodi et al.'s conditional lower bound for two-dimensional range diameter~\cite{dsw12}), and along the way, we introduce a new variant of colored range searching, \emph{color uniqueness query}, which may be of independent interest.

\section{Preliminaries} \label{sec-notation}
The first two results we need are the well-known partition lemma and cutting lemma, both of which are extensively used for solving range-search problems.
\begin{lemma} \label{lem-partition}
\textnormal{\bf (Partition lemma \cite{matouvsek1992efficient})}
Given a set $S$ of $n$ points in $\mathbb{R}^d$ and a parameter $1 \leq r \leq n^{1-\delta}$ for an arbitrarily small constant $\delta>0$, one can compute in $O(n \log n)$ time a partition $\{S_1,\dots,S_r\}$ of $S$ and $r$ simplices $\Delta_1,\dots,\Delta_r$ in $\mathbb{R}^d$ such that \textnormal{\bf (1)} $S_i \subseteq \Delta_i$ for all $i \in \{1,\dots,r\}$, \textnormal{\bf (2)} $|S_i| = O(n/r)$ for all $i \in \{1,\dots,r\}$, and \textnormal{\bf (3)} any hyperplane in $\mathbb{R}^d$ crosses $O(r^{1-1/d})$ simplices among $\Delta_1,\dots,\Delta_r$.
\end{lemma}
\begin{lemma} \label{lem-cutting}
\textnormal{\bf (Cutting lemma \cite{chazelle1993cutting})}
Given a set $\mathcal{H}$ of $n$ hyperplanes in $\mathbb{R}^d$ and a parameter $1 \leq r \leq n$, one can compute in $O(n r^{d-1})$ time a cutting of $\mathbb{R}^d$ into $O(r^d)$ cells each of which is a constant-complexity polytope intersecting $O(n/r)$ hyperplanes in $\mathcal{H}$.
In addition, the algorithm for computing the cutting stores the cells into an $O(r^d)$-space data structure which can report in $O(\log r)$ time, for a specified point in $x \in \mathbb{R}^d$, the cell containing $x$.
\end{lemma}
We shall also use the standard range-reporting data structures for orthogonal, simplex, and halfspace queries, stated in the following lemma: 

\begin{lemma} \label{lem-XRR} 
Given a set $S$ of $n$ points in $\mathbb{R}^d$, one can build in $O(n \log^{O(1)} n)$ time an $O(n \log^{O(1)} n)$-space data structure which can
\begin{enumerate}
  \item[\rm(a)] \label{lem-ORR}
  {\bf (Orthogonal range reporting \cite{berg2008computational})}
 report, for a specified orthogonal box $B$ in $\mathbb{R}^d$, the points in $S \cap B$ in $O(\log^{O(1)} n+k)$ time where $k = |S \cap B|$;
  \item[\rm(b)] \label{lem-SRR}
  {\bf (Simplex range reporting \cite{matouvsek1992efficient})}
  report, for a specified simplex $\Delta$ in $\mathbb{R}^d$, the points in $S \cap \Delta$ in $O(n^{1-1/d} \log^{O(1)} n+k)$ time where $k = |S \cap \Delta|$;
  \item[\rm(c)] \label{lem-HRR}
  {\bf (Halfspace range reporting \cite{matousek1992reporting})}
  report, for a specified halfspace $H$ in $\mathbb{R}^d$, the points in $S \cap H$ in $O(n^{1-1/\lfloor d/2 \rfloor} \log^{O(1)} n+k)$ query time where $k = |S \cap H|$.
\end{enumerate}
\end{lemma}

\noindent
Using a multi-level data structure that combines range trees with the above structures, we can obtain range-reporting structures for query ranges that are the intersections of an orthogonal box and a simplex/halfspace.

\begin{lemma} \label{lem-BXRR}
Given a set $S$ of $n$ points in $\mathbb{R}^d$, one can build in $O(n \log^{O(1)} n)$ time an $O(n \log^{O(1)} n)$-space data structure which can 
\begin{enumerate}
    \item[\rm(a)] \label{lem-BSRR}
    {\bf (Box-simplex range reporting)} report, for a specified orthogonal box $B$ and simplex $\Delta$ in $\mathbb{R}^d$, the points in $S \cap B \cap \Delta$ in $O(\log^{O(1)} n + m^{1-1/d} \log^{O(1)} n+k)$ time where $m = |S \cap B|$ and $k = |S \cap B \cap \Delta|$;
    \item[\rm(b)] \label{lem-BHRR}
    {\bf (Box-halfspace range reporting)} report, for a specified orthogonal box $B$ and halfspace $H$ in $\mathbb{R}^d$, the points in $S \cap B \cap H$ in $O(\log^{O(1)} n + m^{1-1/\lfloor d/2 \rfloor} \log^{O(1)} n+k)$ time where $m = |S \cap B|$ and $k = |S \cap B \cap H|$.
\end{enumerate}
\end{lemma}
\noindent
\textit{Proof.}
We first prove \textbf{(a)}.
The data structure is simply a $d$-dimensional range tree $\mathcal{T}$ built on $S$ in which each node $\mathbf{v} \in \mathcal{T}$ is associated with a simplex range-reporting data structure built on the canonical subset $S(\mathbf{v})$ of $\mathbf{v}$ (Lemma~\ref{lem-SRR}(b)).
The range tree can be built in $O(n \log^{O(1)} n)$ time \cite{berg2008computational}, and the data structure associated to a node $\mathbf{v} \in \mathcal{T}$ can be built in $O(|S(\mathbf{v})| \log^{O(1)} |S(\mathbf{v})|)$ time and occupies $O(|S(\mathbf{v})| \log^{O(1)} |S(\mathbf{v})|)$ space by Lemma~\ref{lem-SRR}(b).
Since $\sum_{\mathbf{v} \in \mathcal{T}} |S(\mathbf{v})| = O(n \log^{O(1)} n)$, we see that the entire data structure can be built in $O(n \log^{O(1)} n)$ time and occupies $O(n \log^{O(1)} n)$ space.
To answer a box-simplex range-reporting query $(B,\Delta)$, we first find the $O(\log^d n)$ canonical nodes $\mathbf{v}_1,\dots,\mathbf{v}_t$ in $\mathcal{T}$ corresponding to the box $B$, which takes $O(\log^d n)$ time \cite{berg2008computational}.
We have $S \cap B = \bigcup_{i=1}^t S(\mathbf{v}_i)$ and $S(\mathbf{v}_i) \cap S(\mathbf{v}_j) = \emptyset$ if $i \neq j$. 
Therefore, $m = |S \cap B| = \sum_{i=1}^t |S(\mathbf{v}_i)|$.
Then for each $i \in \{1,\dots,t\}$, we use the simplex range-reporting data structure associated to $\mathbf{v}_i$ to report the points in $S(\mathbf{v}_i) \cap \Delta$, taking $O(|S(\mathbf{v}_i)|^{1-1/d} \log^{O(1)} |S(\mathbf{v}_i)| + k_i)$ time where $k_i = |S(\mathbf{v}_i) \cap \Delta|$, by Lemma~\ref{lem-SRR}(b).
Since $t = O(\log^d n)$ and $\sum_{i=1}^t k_i = |S \cap B \cap \Delta| = k$, the total query time is $O(\log^{O(1)} n + m^{1-1/d} \log^{O(1)} n+k)$.

We then prove \textbf{(b)}.
The data structure is simply a $d$-dimensional range tree $\mathcal{T}$ built on $S$ in which each node $\mathbf{v} \in \mathcal{T}$ is associated with a halfspace range-reporting data structure built on the canonical subset $S(\mathbf{v})$ of $\mathbf{v}$ (Lemma~\ref{lem-HRR}(c)).
The range tree can be built in $O(n \log^{O(1)} n)$ time \cite{berg2008computational}, and the data structure associated to a node $\mathbf{v} \in \mathcal{T}$ can be built in $O(|S(\mathbf{v})| \log |S(\mathbf{v})|)$ time and occupies $O(|S(\mathbf{v})|)$ space by Lemma~\ref{lem-HRR}(c).
Since $\sum_{\mathbf{v} \in \mathcal{T}} |S(\mathbf{v})| = O(n \log^{O(1)} n)$, we see that the entire data structure can be built in $O(n \log^{O(1)} n)$ time and occupies $O(n \log^{O(1)} n)$ space.
To answer a box-halfspace range-reporting query $(B,H)$, we first find the $O(\log^d n)$ canonical nodes $\mathbf{v}_1,\dots,\mathbf{v}_t$ in $\mathcal{T}$ corresponding to the box $B$, which takes $O(\log^d n)$ time \cite{berg2008computational}.
We have $S \cap B = \bigcup_{i=1}^t S(\mathbf{v}_i)$ and $S(\mathbf{v}_i) \cap S(\mathbf{v}_j) = \emptyset$ if $i \neq j$. 
Therefore, $m = |S \cap B| = \sum_{i=1}^t |S(\mathbf{v}_i)|$.
Then for each $i \in \{1,\dots,t\}$, we use the halfspace range-reporting data structure associated to $\mathbf{v}_i$ to report the points in $S(\mathbf{v}_i) \cap \Delta$, taking $O(|S(\mathbf{v}_i)|^{1-1/\lfloor d/2 \rfloor} \log^{O(1)} |S(\mathbf{v}_i)| + k_i)$ time where $k_i = |S(\mathbf{v}_i) \cap H|$.
Since $t = O(\log^d n)$ and $\sum_{i=1}^t k_i = |S \cap B \cap H| = k$, the total query time is $O(\log^{O(1)} n + m^{1-1/\lfloor d/2 \rfloor} \log^{O(1)} n+k)$.
\hfill $\Box$


\section{Orthogonal RCP queries} \label{sec-orthogonal}

\subsection{Data structure}
Let $S$ be a set of $n$ points in $\mathbb{R}^d$.
In this section, we show how to build a RCP data structure on $S$ for orthogonal queries.
First, we build a (standard) $d$-dimensional range tree $\mathcal{T}$ on $S$.
Each node $\mathbf{u}$ of $\mathcal{T}$ corresponds to a \textit{canonical subset} of $S$, which we denote by $S(\mathbf{u})$.
We say $\mathbf{u}$ is a \textit{heavy} node if $|S(\mathbf{u})| \geq \sqrt{n}$.
For every pair $(\mathbf{u},\mathbf{v})$ of heavy nodes, we compute the closest pair $\phi_{\mathbf{u},\mathbf{v}}$ in $S(\mathbf{u}) \cup S(\mathbf{v})$; denote by $\varPhi$ the set of all these pairs.
Then we build an orthogonal range-reporting data structure $\mathcal{D}(S)$ on $S$ (Lemma~\ref{lem-ORR}(a)).
Our orthogonal RCP data structure consists of the range tree $\mathcal{T}$, the data structure $\mathcal{D}(S)$, and the pair set $\varPhi$.
\smallskip

\begin{figure}
    \centering
    \includegraphics[scale=0.8]{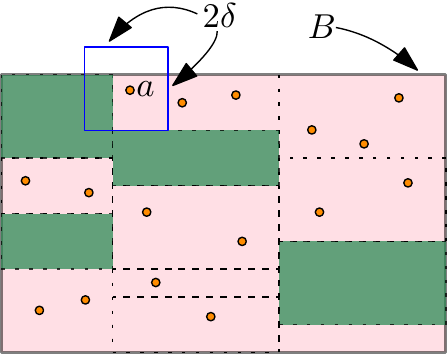}\hspace{6em}
    \includegraphics[scale=0.7]{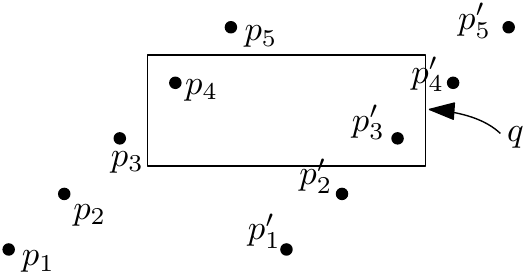}
    \caption{(Left) The canonical nodes in the range tree 
    ${\cal T}$ break the query box $B$ into thirteen disjoint regions. The 
    green regions correspond to set $I$ (the heavy nodes). The orange points form the set $L$. For one of the points in 
    $L$ (denoted by $a$), the box $\Box_a$ is shown in blue. The crucial property is that the number of points which lie in $B\cap \Box_a$ is $O(1)$.
    (Right) Reduction from the set intersection query to the color uniqueness query. The set intersection query is to test if $S_4$ and $S_3$ are disjoint, and the query rectangle $q$ for the color uniqueness query exactly contains points 
    $p_4$ and $p'_3$.
    }
    \label{fig:rectangle}\label{fig:SI-to-CU}
\end{figure}

\noindent
\textbf{Query procedure.}
Consider a query box $B$ in $\mathbb{R}^d$.
Our goal is to find the closest pair in $S \cap B$ using the data structure described above.
By searching in the range tree $\mathcal{T}$, we can find $t = O(\log^{O(1)} n)$ canonical nodes $\mathbf{c}_1,\dots,\mathbf{c}_t$ corresponding to $B$.
We have $S \cap B = \bigcup_{i=1}^t S(\mathbf{c}_i)$.
Let $I = \{i: \mathbf{c}_i \text{ is a heavy node}\}$ and $I' = \{1,\dots,t\} \backslash I$.  (See Figure~\ref{fig:rectangle}(left).)
For all $i,j \in I$, we obtain the pair $\phi_{\mathbf{c}_i,\mathbf{c}_j}$ from $\varPhi$ and take the closest one $\phi \in \{\phi_{\mathbf{c}_i,\mathbf{c}_j}: i,j \in I\}$.
On the other hand, we compute $L = \bigcup_{i \in I'} S(\mathbf{c}_i)$.
We take the closest pair $\phi'$ in $L$.
Let $\delta = \min\{|\phi|,|\phi'|\}$.
For each $a \in L$, let $\Box_a$ be the hypercube centered at $a$ with side-length $2\delta$.
We query, for each $a \in L$, the box range-reporting data structure $\mathcal{D}(S)$ with $\Box_a \cap B$ to obtain the set $P_a = S \cap \Box_a \cap B$.
After this, for each $a \in L$, we compute a pair $\psi_a$ consisting of $a$ and the nearest neighbor of $a$ in $P_a \backslash \{a\}$.
We then take the closest one $\psi \in \{\psi_a: a \in L\}$.
Finally, if $|\psi| < |\phi|$, then we return $\psi$ as the answer; otherwise, we return $\phi$ as the answer.

We now verify the correctness of the above query procedure.
Let $\phi^* = (a,b)$ be the closest pair in $S \cap \Delta$.
It suffices to show that $|\phi| \leq |\phi^*|$ or $|\psi| \leq |\phi^*|$.
Suppose $a \in S(\mathbf{c}_i)$ and $b \in S(\mathbf{c}_j)$.
If $i,j \in I$, then $|\phi| \leq |\phi_{\mathbf{c}_i,\mathbf{c}_j}| \leq |\phi^*|$ and we are done.
Otherwise, either $i \in I'$ or $j \in I'$; assume $i \in I'$ without loss of generality.
It follows that $a \in L$.
Since $\phi^*$ is the closest pair in $S \cap B$, we have $|\phi^*| \leq |\phi|$ and $|\phi^*| \leq |\phi'|$, which implies that the distance between $a$ and $b$ is at most $\delta$.
Therefore, $b \in P_a$.
Now we have $|\psi| \leq |\psi_a| \leq |\phi^*|$, which completes the proof of the correctness.

\smallskip
\noindent
\textbf{Analysis.}
We analyze the performance (space, query time, and preprocessing time) of our orthogonal RCP data structure.
To this end, we first bound the number of the heavy nodes.
The lemma below follows immediately from the well-known fact that the sum of sizes of the canonical subsets in a range tree is $O(n\log^d n)$.

\begin{lemma}\label{lem-heavy}
There are $O(\sqrt{n}\log^{O(1)}n)$ 
heavy nodes in $\mathcal{T}$.
\end{lemma}
\newcommand{\PROOFHEAVY}{
For convenience, we call a node in a range tree $\alpha$-\textit{heavy} if its canonical subset is of size at least $\alpha$.
Then the heavy nodes defined before are just $\sqrt{n}$-heavy nodes.
Let $f(k_1,k_2,d)$ be the number of the $2^{k_2}$-heavy nodes in a $d$-dimensional range tree built on $2^{k_1}$ points.
We claim that $f(k_1,k_2,d) = O(2^{k_1-k_2} \cdot k_1^{d-1})$ for any constant $d \geq 1$.
We use induction on the dimension $d$.
In a 1-dimensional range tree built on $2^{k_1}$ points, the number of the $2^{k_2}$-heavy nodes is clearly $O(2^{k_1-k_2})$.
Suppose the claim holds for range trees of dimension $1,\dots,d-1$, and we consider the $d$-dimensional case.
Let $\mathcal{T}$ be a $d$-dimensional range tree built on a set $S$ of $2^{k_1}$ points.
By definition, $\mathcal{T}$ is essentially a 1-dimensional range tree (called \textit{main tree}) built on $S$ in which each node is associated with a $(d-1)$-dimensional range tree built on its canonical subset.
The number of the $2^{k_2}$-heavy nodes in the main tree is $O(2^{k_1-k_2})$.
In the main tree, there are $2^{k_1-i}$ nodes whose canonical subsets are of size $2^i$, and the number of the $2^{k_2}$-heavy nodes in the $(d-1)$-dimensional range tree associated with each of these nodes is $f(i,k_2,d-1)$.
So we have the recurrence
\begin{equation*}
    f(k_1,k_2,d) = O(2^{k_1-k_2}) + \sum_{i=0}^{k_1} 2^{k_1-i} \cdot f(i,k_2,d-1).
\end{equation*}
By our induction hypothesis, $f(i,k_2,d-1) = O(2^{i-k_2} \cdot i^{d-2})$.
Therefore, using the above recurrence, we can deduce that $f(k_1,k_2,d) = O(2^{k_1-k_2} \cdot k_1^{d-1})$.
With this claim in hand, we can now prove the lemma.
Without loss of generality, we may assume $n = 4^k$ and hence $\sqrt{n} = 2^k$.
Since $f(2k,k,d) = O(2^k \log^{d-1} (2k))$, there are $O(\sqrt{n} \log^{d-1} n)$ $\sqrt{n}$-heavy nodes.
\hfill $\Box$
\smallskip
}

\noindent
By the above lemma, the space of the data structure is $O(n \log^{O(1)} n)$.
Indeed, the range tree $\mathcal{T}$ and the data structure $\mathcal{D}(S)$ both occupy $O(n \log^{d-1} n)$ space, and the pair-set $\varPhi$ takes $O(n \log^{2d-2} n)$ space as there are $O(\sqrt{n}\log^{O(1)}n)$ 
heavy nodes.
The preprocessing time is $O(n \sqrt{n} \log^{O(1)} n)$.
Indeed, building the range tree $\mathcal{T}$ and the data structure $\mathcal{D}(S)$ takes $O(n \log^{O(1)} n)$ time.
We claim that the pair-set $\varPhi$ can be computed in $O(n \sqrt{n} \log^{O(1)} n)$ time.
We first find the set $\mathcal{H}$ of heavy nodes, which can be done in $O(n \log^{O(1)} n)$ time by simply checking every node of $\mathcal{T}$.
For two pairs $(\mathbf{u},\mathbf{v})$ and $(\mathbf{u}',\mathbf{v}')$ of nodes in $\mathcal{H}$, we write $(\mathbf{u},\mathbf{v}) \preceq (\mathbf{u}',\mathbf{v}')$ if $|S(\mathbf{u})|+|S(\mathbf{v})| \leq |S(\mathbf{u}')|+|S(\mathbf{v}')|$.
Then ``$\preceq$'' is a partial order on $\mathcal{H} \times \mathcal{H}$.
We consider the pairs of heavy nodes in this partial order from the smallest to the greatest.
For each pair $(\mathbf{u},\mathbf{v})$, we compute $\phi_{\mathbf{u},\mathbf{v}}$ as follows.
If $|S(\mathbf{u})| < 2 \sqrt{n}$ and $|S(\mathbf{v})| < 2 \sqrt{n}$, we explicitly compute $S(\mathbf{u}) \cup S(\mathbf{v})$ and then compute $\phi_{\mathbf{u},\mathbf{v}}$ using the standard closest-pair algorithm in $O(\sqrt{n} \log n)$ time.
Otherwise, either $|S(\mathbf{u})| \geq 2 \sqrt{n}$ or $|S(\mathbf{v})| \geq 2 \sqrt{n}$.
Without loss of generality, assume $|S(\mathbf{u})| \geq 2 \sqrt{n}$.
Then the two children $\mathbf{u}_1$ and $\mathbf{u}_2$ of $\mathbf{u})$ are both heavy.
Note that $\phi_{\mathbf{u},\mathbf{v}}$ is the closest one among $\phi_{\mathbf{u}_1,\mathbf{v}},\phi_{\mathbf{u}_2,\mathbf{v}},\phi_{\mathbf{u}_1,\mathbf{u}_2}$ by construction.
Also note that $(\mathbf{u}_1,\mathbf{v}) \preceq (\mathbf{u},\mathbf{v})$, $(\mathbf{u}_2,\mathbf{v}) \preceq (\mathbf{u},\mathbf{v})$, $(\mathbf{u}_1,\mathbf{u}_2) \preceq (\mathbf{u},\mathbf{v})$, thus $\phi_{\mathbf{u}_1,\mathbf{v}},\phi_{\mathbf{u}_2,\mathbf{v}},\phi_{\mathbf{u}_1,\mathbf{u}_2}$ have already been computed when considering $(\mathbf{u},\mathbf{v})$.
With $\phi_{\mathbf{u}_1,\mathbf{v}},\phi_{\mathbf{u}_2,\mathbf{v}},\phi_{\mathbf{u}_1,\mathbf{u}_2}$ in hand, we can compute $\phi_{\mathbf{u},\mathbf{v}}$ in $O(1)$ time.
In sum, $\phi_{\mathbf{u},\mathbf{v}}$ can be computed in $O(\sqrt{n} \log n)$ time in any case.
Since $|\mathcal{H} \times \mathcal{H}| = O(n \log^{O(1)} n)$, we can compute $\varPhi$ in $O(n \sqrt{n} \log^{O(1)} n)$ time.
This completes the discussion of the preprocessing time.
Next, we analyze the query time.
Finding the canonical nodes $\mathbf{c}_1,\dots,\mathbf{c}_t$ takes $O(\log^{O(1)} n)$ time, so does computing the index sets $I$ and $I'$.
Obtaining the set $\{\phi_{\mathbf{c}_i,\mathbf{c}_j}: i,j \in I\}$ and computing $\phi$ takes $O(\log^{O(1)} n)$ time since $|I| \leq t$ and $t = O(\log^{O(1)} n)$.
Computing $\phi'$ requires $O(\sqrt{n} \log^{O(1)} n)$ time, because $|L| = O(t \sqrt{n}) = O(\sqrt{n} \log^{O(1)} n)$.
For a point $a \in L$, reporting the points in $P_a$ takes $O(\log^{O(1)}n + |P_a|)$ time.
Therefore, computing all the $P_a$'s can be done in $O(|L| \log^{O(1)}n + \sum_{a \in L}|P_a|)$ time.
To bound this quantity, we observe the following fact.
\begin{lemma}
$|P_a| = O(1)$ for all $a \in L$.
\end{lemma}
\textit{Proof.}
We have $S \cap B = (\bigcup_{i \in I} S(\mathbf{u}_i)) \cup L$.
It suffices to show that $|(\bigcup_{i \in I}S(\mathbf{u}_i)) \cap \Box_a| = O(1)$ and $|L \cap \Box| = O(1)$.
Both facts follow from the Pigeonhole Principle readily.
Indeed, we have $|(\bigcup_{i \in I}S(\mathbf{u}_i)) \cap \Box_a| = O(1)$ because $\phi$ is the closest pair in $\bigcup_{i \in I}S(\mathbf{u}_i)$ and $|\phi| \geq \delta$.
We have $|L \cap \Box| = O(1)$ because $\phi'$ is the closest pair in $L$ and $|\phi'| \geq \delta$.
This completes the proof.
\hfill $\Box$
\smallskip

\noindent
By the above lemma and the fact $|L| = O(\sqrt{n} \log^{O(1)} n)$, we can compute all the $P_a$'s in $O(\sqrt{n} \log^{O(1)} n)$ time.
The pair $\psi$ can be directly obtained after knowing all the $P_a$'s, hence the total query time is $O(\sqrt{n} \log^{O(1)} n)$.
We conclude the following.
\begin{theorem}\label{thm:orthogonal-ub}
Given a set $S$ of $n$ points in $\mathbb{R}^d$, one can construct in $\tO(n\sqrt{n})$ time an orthogonal RCP data structure on $S$ with $\tO(n)$ space and $\tO(\sqrt{n})$ query time.
\end{theorem}


\subsection{Conditional hardness}
In this subsection, we prove a 
conditional lower-bound for the orthogonal RCP query, which shows that the upper bound given in Theorem~\ref{thm:orthogonal-ub} is tight, ignoring $\log n$ factors.
following lower-bound matches the upper-bound of Theorem~\ref{thm:orthogonal-ub}.
First, we define the following problem~\cite{pr14}.

\begin{problem} {\bf (Set intersection query)}
The input is a collection of 
sets $S_1, S_2,\ldots,S_m$ of positive reals 
such that $\sum_{i=1}^{m}|S_i|=n$. Given query 
indices $i$ and $j$, report if $S_i$ and $S_j$ 
are disjoint, or not?
\end{problem}

This problem can be viewed as a query version of
Boolean matrix multiplication, and is conjectured to be {\em hard}: 
in the cell-probe model without the floor function
and where the cardinality of each set $S_i$ is 
upper-bounded by $\log^{O(1)}m$, any data structure
to answer the set intersection problem in 
$\tilde{O}(\alpha)$ time requires 
$\tilde{\Omega}((n/\alpha)^2)$ space, 
for $1\leq \alpha \leq n$~\cite{dsw12,pr14}.
In particular, any linear-space structure
is believed to require $\tilde{\Omega}(\sqrt{n})$ time.

Next we introduce an intermediate geometric problem,
which may be of independent interest:

\begin{problem} {\bf (Set intersection query)}
The input is a set $S$ of $n$ {\em colored} 
points in $\IR^2$. Specifically, 
let $C$ be a collection of distinct colors,
and each point $p\in S$ is associated with some
color from $C$. Given a query rectangle $q$, 
report if all the colors are unique in $S\cap q$? 
In other words, is there a color which has at least 
two points in $S\cap q$?
\end{problem}

We will perform a two-step reduction: first, 
reduce the set intersection query to the 
color uniqueness query, and then reduce 
the two-dimensional color uniqueness query to the three-dimensional  orthogonal 
RCP query.

\smallskip
\noindent
\textbf{Reduction from set intersection to color 
uniqueness in $\IR^2$.} Given an instance of the set intersection 
query, we will construct an instance of the color 
uniqueness query. Let $p_1=(1,1), p_2=(2,2),\ldots, 
p_{m}=(m,m)$, and $p_1'=(m+1,1), 
p_2'=(m+2,2),\ldots, p_{m}'=(2m,m)$.
Next, assign a unique color to each distinct 
element in $S_1 \cup S_2 \cup \ldots \cup S_m$. 
Now replace each point $p_i$  with 
$|S_i|$ {\em new} points such that (a) the new points are within 
a distance of $\varepsilon \ll 1$ from $p_i$, and 
(b)  each new point picks a distinct color from 
the colors  assigned to the  
elements in $S_i$. 
Perform a similar operation for points $p_i'$.
Let $P$ be the collection of these $2n$ new points.

To answer if $S_i$ and $S_j$ are disjoint ($j< i$), 
we ask a color uniqueness query on $P$ with an axis-aligned rectangle 
$q=[i-\varepsilon, m+j+\varepsilon]\times [j-\varepsilon, 
i+\varepsilon]$ (see Figure~\ref{fig:SI-to-CU}(right)). 
If there is a color which contains two 
points, then we report that $S_i$ and $S_j$ are not disjoint; 
otherwise, we report that $S_i$ and $S_j$ are disjoint.
The correctness is easy to see: the key observation is that
$q$ exactly contains the points of 
$S_i$ and $S_j$. Therefore, $S_i$ and $S_j$ are disjoint 
iff all the colors are unique in $P\cap q$.
Reductions of this flavor have been performed 
before~\cite{akss18,dsw12,krsv07,rj12}.


\newcommand{\dmax}{d_{\mbox{\scriptsize max}}}
\smallskip
\noindent
\textbf{Reduction from color uniqueness in $\IR^2$ to 
orthogonal RCP in $\IR^3$.} Given an instance of the color 
uniqueness query, we will now construct an instance of 
the orthogonal RCP query in $\IR^3$. Let $\dmax$ be 
the maximum Euclidean distance between any two points in 
$S$, and let $c_1,c_2,\ldots,c_{|C|}$ be the $|C|$ colors 
in the dataset. Then each point $p=(p_x,p_y)\in S$ with color $c_i$ 
is mapped to a 3-d point $p'=(p_x,p_y, 2\cdot i\cdot \dmax)$. 
Let $P$ be the collection of these $n$ newly mapped points.

To answer the color uniqueness query for a rectangle $q$, 
we will ask an orthogonal RCP query on $P$ with the query 
box $q\times (-\infty, \infty)$. If the closest-pair distance 
is less than or equal to $\dmax$, then we report that there 
is a color which contains at least two points inside $q$; 
otherwise, we report that all the colors are unique inside $q$.
Once again, the correctness is easy to see: the key observation  
is that the distance between points of different colors in $P$
is at least $2\cdot \dmax$.

The above two reductions together implies our conditional lower bound, which is presented in the following theorem.

\begin{theorem}\label{thm:orthogonal-lb}
The orthogonal RCP query is at least 
as hard as the set intersection query.
\end{theorem}

\section{Simplex RCP queries} \label{sec-simplex}
Let $S$ be a set of $n$ points in $\mathbb{R}^d$, and $r$ be a parameter to be specified shortly.
In this section, we show how to build a RCP data structure on $S$ for simplex queries.
First, we use Lemma~\ref{lem-partition} to compute a partition $\{S_1,\dots,S_r\}$ of $S$ and $r$ simplices $\Delta_1,\dots,\Delta_r$ in $\mathbb{R}^d$ satisfying the conditions in the lemma.
For every $i,j \in \{1,\dots,r\}$, we compute the closest pair $\phi_{i,j}$ in $S_i \cup S_j$; denote by $\varPhi$ the set of all these pairs.
Then we build a box-simplex range-reporting data structure $\mathcal{D}'(S)$ on $S$ (Lemma~\ref{lem-BSRR}(a)).
Our simplex RCP data structure consists of the partition $\{S_1,\dots,S_r\}$, the simplices $\Delta_1,\dots,\Delta_r$, the data structure $\mathcal{D}'(S)$, and the pair set~$\varPhi$.
\smallskip

\begin{figure}
    \centering
    \includegraphics[scale=0.8]{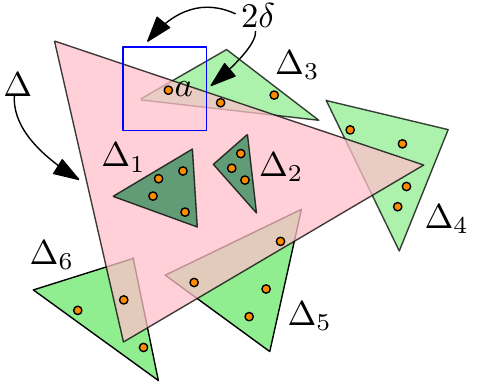}
    \caption{$I=\{\Delta_1, \Delta_2\}$ and 
    $I'=\{\Delta_3,\Delta_4,\Delta_5, \Delta_6 \}$.}
    \label{fig:simplex}
\end{figure}

\noindent
\textbf{Query procedure.}
Consider a query simplex $\Delta$ in $\mathbb{R}^d$.
Our goal is to find the closest pair in $S \cap \Delta$ using the data structure described above.
We first compute two index sets $I = \{i: \Delta_i \subseteq \Delta\}$, $I' = \{i: \Delta_i \nsubseteq \Delta \text{ and } \Delta_i \cap \Delta \neq \emptyset\}$.
(See Figure~\ref{fig:simplex}.)
These index sets are computed by explicitly considering the $r$ simplices $\Delta_1,\dots,\Delta_r$.
For all $i,j \in I$, we obtain the pair $\phi_{i,j}$ from $\varPhi$ and take the closest one $\phi \in \{\phi_{i,j}: i,j \in I\}$.
On the other hand, we compute a set $L = (\bigcup_{i \in I'} S_i) \cap \Delta$ by simply checking, for every $i \in I'$ and every $a \in S_i$, whether $a \in \Delta$.
We take the closest pair $\phi'$ in $L$.
Let $\delta = \min\{|\phi|,|\phi'|\}$.
For each $a \in L$, let $\Box_a$ be the hypercube centered at $a$ with side length $2\delta$.
We query, for each $a \in L$, the box-simplex range-reporting data structure $\mathcal{D}'(S)$ with $\Box_a$ and $\Delta$ to obtain the set $P_a = S \cap \Box_a \cap \Delta$.
After this, for each $a \in L$, we compute a pair $\psi_a$ consisting of $a$ and the nearest neighbor of $a$ in $P_a \backslash \{a\}$.
We then take the closest one $\psi \in \{\psi_a: a \in L\}$.
Finally, if $|\psi| < |\phi|$, then we return $\psi$ as the answer; otherwise, we return $\phi$ as the answer.

We now verify the correctness of the above query procedure.
Let $\phi^* = (a,b)$ be the closest pair in $S \cap \Delta$.
It suffices to show that $|\phi| \leq |\phi^*|$ or $|\psi| \leq |\phi^*|$.
Suppose $a \in S_i$ and $b \in S_j$.
We first notice that $i,j \in I \cup I'$.
Indeed, if $i \notin I \cup I'$ (resp., $j \notin I \cup I'$), then $\Delta_i \cap \Delta = \emptyset$ (resp., $\Delta_j \cap \Delta = \emptyset$) and hence $S_i \cap \Delta = \emptyset$ (resp., $S_j \cap \Delta = \emptyset$), which contradicts the fact that $a \in S_i \cap \Delta$ (resp., $b \in S_i \cap \Delta$).
If $i,j \in I$, then $|\phi| \leq |\phi_{i,j}| \leq |\phi^*|$ and we are done.
Otherwise, either $i \in I'$ or $j \in I'$; assume $i \in I'$ without loss of generality.
It follows that $a \in L$.
Since $\phi^*$ is the closest pair in $S \cap \Delta$, we have $|\phi^*| \leq |\phi|$ and $|\phi^*| \leq |\phi'|$, which implies that the distance between $a$ and $b$ is at most $\delta$.
Therefore, $b \in P_a$.
Now we have $|\psi| \leq |\psi_a| \leq |\phi^*|$, which completes the proof of the correctness.
\smallskip

\noindent
\textbf{Analysis.}
We analyze the performance (space, query time, and preprocessing time) of our simplex RCP data structure.
The space of the data structure is $O(n \log^{O(1)} n + r^2)$, because $\mathcal{D}'(S)$ occupies $O(n \log^{O(1)} n)$ space and $\varPhi$ occupies $O(r^2)$ space.
The preprocessing time is $O(nr \log^{O(1)} n)$.
Indeed, computing the partition $\{S_1,\dots,S_r\}$ and the simplices $\Delta_1,\dots,\Delta_r$ takes $O(n \log n)$ time by Lemma~\ref{lem-partition}.
Computing $\phi_{i,j}$ for some fixed $i,j \in \{1,\dots,r\}$ can be done in $O((n/r) \log (n/r))$ time using the standard closest-pair algorithm, because $|S_i \cup S_j| = O(n/r)$.
It follows that computing $\varPhi$ takes $O(nr \log n)$ time.
Finally, building the data structure $\mathcal{D}'(S)$ requires $O(n \log^{O(1)} n)$ time.
As such, our simplex RCP data structure can be constructed in $O(nr \log^{O(1)} n)$ time.
Next, we analyze the query time.
The index sets $I$ and $I'$ are computed in $O(r)$ time.
Obtaining the set $\{\phi_{i,j}: i,j \in I\}$ and computing $\phi$ requires $O(r^2)$ time.
The set $L$ is computed by explicitly considering all the points in $\bigcup_{i \in I'} S_i$ in $O(\sum_{i \in I'} |S_i|)$ time.
We notice that $|I'| = O(r^{1-1/d})$, since each facet of $\Delta$ only intersects $O(r^{1-1/d})$ simplices among $\Delta_1,\dots,\Delta_r$ by Lemma~\ref{lem-partition}.
It follows that $\sum_{i \in I'} |S_i| = O(n/r^{1/d})$, because $|S_i| = O(n/r)$.
That says, $L$ can be computed in $O(n/r^{1/d})$ time and in particular, $|L| = O(n/r^{1/d})$.
Once $L$ is obtained, $\phi'$ can be computed in $O((n/r^{1/d})\log(n/r^{1/d}))$ time using the standard closest-pair algorithm.
For a point $a \in L$, reporting the points in $P_a$ takes $O(\log^{O(1)} n + m_a^{1-1/d} \log^{O(1)} m_a + |P_a|)$ time where $m_a = |S \cap \Box_a|$, by Lemma~\ref{lem-BSRR}(a).
Therefore, computing all the $P_a$'s can be done in $O(\sum_{a \in L} m_a^{1-1/d} \log^{O(1)} n + \sum_{a \in L} |P_a|)$ time.
To bound this quantity, we observe the following fact.
\begin{lemma} \label{lem-number}
$\sum_{a \in L} m_a = O(n)$ and $|P_a| = O(1)$ for all $a \in L$.
\end{lemma}
\textit{Proof.}
We first prove $\sum_{a \in L} m_a = O(n)$.
Consider a point $p \in S$.
Let $\Box_p$ be the hypercube centered at $p$ with side-length $2\delta$.
Note that $p \in P_a$ only if $a \in \Box_p$ for all $a \in L$.
Since $\phi'$ is the closest pair in $L$ and $|\phi'| \geq \delta$, we have $L \cap \Box_p = O(1)$ by the Pigeonhole Principle.
Therefore, only a constant number of points in $L$ is contained in $p$.
In other words, any point $p \in S$ is contained in $P_a$ for only a constant number of $a \in L$, which implies $\sum_{a \in L} m_a = O(n)$.
Next, we prove that $|P_a| = O(1)$ for all $a \in L$.
Clearly, $S \cap \Delta = (\bigcup_{i \in I} S_i) \cup L$.
So it suffices to show that $|(\bigcup_{i \in I} S_i) \cap \Box_a| = O(1)$ and $|L \cap \Box_a| = O(1)$.
Both facts follow from the Pigeonhole Principle readily.
Indeed, we have $|(\bigcup_{i \in I} S_i) \cap \Box_a| = O(1)$ because $\phi$ is the closest pair in $\bigcup_{i \in I} S_i$ and $|\phi| \geq \delta$.
We have $|L \cap \Box_a| = O(1)$ because $\phi'$ is the closest pair in $L$ and $|\phi'| \geq \delta$.
This completes the proof of $|P_a| = O(1)$.
\hfill $\Box$
\smallskip

\noindent
By the above lemma and H\"older's inequality, we have
\begin{equation*}
    \sum_{a \in L} m_a^{1-1/d} \leq 
    O(n^{1-1/d} |L|^{1/d}) = O\left(\frac{n}{r^{1/d^2}}\right),
\end{equation*}
which implies that computing all the $P_a$'s takes $O((n \log^{O(1)} n)/r^{1/d^2})$ time.
The pair $\psi$ can be directly obtained after knowing all the $P_a$'s.
Hence, the total query time is $O(r^2 + (n \log^{O(1)} n)/r^{1/d^2})$.
Setting $r = n^{d^2/(2d^2+1)}$ gives: 
\begin{theorem} \label{thm-SRCP}
Given a set $S$ of $n$ points in $\mathbb{R}^d$, one can construct in $\tO(n^{(3d^2+1)/(2d^2+1)})$ time a simplex RCP data structure on $S$ with $\tO(n)$ space and $\tO(n^{1-1/(2d^2)})$ query time.
\end{theorem}

Note that our data structure above can also handle constant-complexity polytope RCP queries (with the same query procedure and query time).
In other words, the data structure can be used to report, for specified $O(1)$ halfspaces $H_1,\dots,H_c$ in $\mathbb{R}^d$, the closest pair in $S \cap (\bigcap_{i=1}^c H_i)$ in $\tO(n^{1-1/(2d^2)})$ time.

\section{Halfspace RCP queries} \label{sec-halfspace}
Let $S$ be a set of $n$ points in $\mathbb{R}^d$, and $r$ be a parameter to be specified shortly.
In this section, we show how to build an RCP data structure on $S$ for halfspace queries.
The same method can also result in an RCP data structure for ball queries, using the standard lifting argument.
Since halfspace query is a special case of simplex query, the simplex RCP data structure in the last section can be directly used to answer halfspace RCP queries.
But in fact, for halfspace RCP queries, we can achieve better bounds.

\begin{figure}
    \centering
    \includegraphics[scale=0.7]{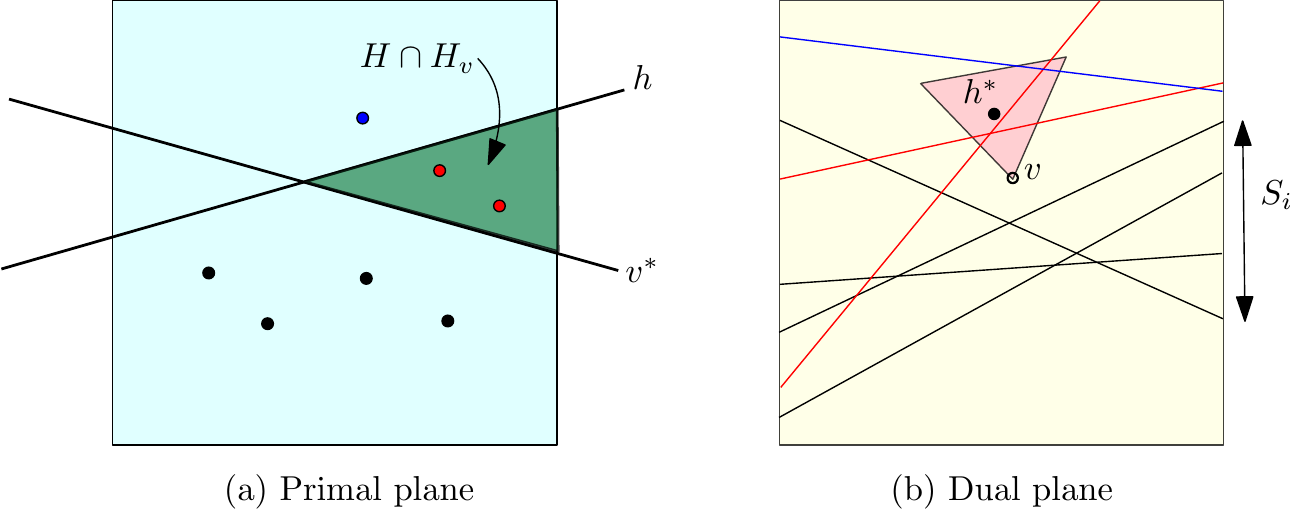}
    \caption{The dataset shown in (a) consists of seven points. 
    The dual $h^{*}$ of the query hyperplane $h$ lies inside the cell $\Xi_i$ shown in pink in (b). The closest pair among the black points, $\phi_i$, is computed in the preprocessing phase itself (since the dual of the black points is the set $S_i$). The red points belong to set $L$ and are explicitly reported during the query procedure.}
    \label{fig:halfspace}
\end{figure}

It suffices to consider the halfspaces which are regions below non-vertical hyperplanes, namely, halfspaces of the form $x_d \leq a_1 x_1 + \cdots + a_{d-1} x_{d-1}$.
By duality, a point $a \in S$ maps to a hyperplane $a^*$ in the dual space (which is also a copy of $\mathbb{R}^d$).
Also, a non-vertical hyperplane $h$ in the primal $\mathbb{R}^d$ maps to a point $h^*$ in the dual space.
The property of duality guarantees that $a$ is above (resp., below) $h$ iff $h^*$ is above (resp., below) $a^*$ for all $a \in S$ and all hyperplanes $h$ (see Figure~\ref{fig:halfspace}).
Define $\mathcal{H} = \{a^*: a \in S\}$.
We use Lemma~\ref{lem-cutting} to cut $\mathbb{R}^d$ (the dual space) into $R = O(r^d)$ cells $\Xi_1,\dots,\Xi_R$ each of which is a constant-complexity polytope intersecting $O(n/r)$ hyperplanes in $\mathcal{H}$.
For $i \in \{1,\dots,R\}$, let $S_i = \{a: a^* \text{ is below } \Xi_i\}$.
We associate to the cell $\Xi_i$ the closest pair $\phi_i$ in $S_i$.
Furthermore, we build a simplex range-reporting data structure $\mathcal{D}(S)$ on $S$ (Lemma~\ref{lem-SRR}(b)) and a box-halfspace range-reporting data structure $\mathcal{D}'(S)$ in $S$ (Lemma~\ref{lem-BHRR}(b)).
Our halfspace RCP data structure consists of the cells $\Xi_1,\dots,\Xi_R$ (with the associated pairs $\phi_1,\dots,\phi_r$) and the data structures $\mathcal{D}(S)$ and $\mathcal{D}'(S)$.
The cells $\Xi_1,\dots,\Xi_R$ are stored in the way mentioned in Lemma~\ref{lem-cutting} (so that we can do point location efficiently).
\smallskip

\noindent
\textbf{Query procedure.}
Consider a query halfspace $H$ that is the region below a non-vertical hyperplane $h$.
Our goal is to find the closest pair in $S \cap H$ using the data structure described above.
To this end, we first find the cell $\Xi_i$ such that $h^* \in \Xi_i$.
Let $V$ be the set of the vertices of $\Xi_i$.
We have $V = O(1)$ by Lemma~\ref{lem-cutting}.
For every $v \in V$, let $H_v$ be the halfspace above the non-vertical hyperplane $v^*$ in the primal $\mathbb{R}^d$.
Using $\mathcal{D}(S)$, we find the points in $S \cap (H \cap H_v)$ for all $v \in V$ and obtain the set $L = \bigcup_{v \in V} S \cap (H \cap H_v)$.
We take the closest pair $\phi'$ in $L$.
Let $\delta = \min\{|\phi_i|,|\phi'|\}$ (recall that $\phi_i$ is the pair associated to $\Xi_i$).
For each $a \in L$, let $\Box_a$ be the hypercube centered at $a$ with side-length $2\delta$.
We query, for each $a \in L$, the box-halfspace range-reporting data structure $\mathcal{D}'(S)$ with $\Box_a$ and $H$ to obtain the set $P_a = S \cap \Box_a \cap H$.
After this, for each $a \in L$, we compute a pair $\psi_a$ consisting of $a$ and the nearest neighbor of $a$ in $P_a \backslash \{a\}$.
We then take the closest one $\psi \in \{\psi_a: a \in L\}$.
Finally, if $|\psi| < |\phi_i|$, then we return $\psi$ as the answer; otherwise, we return $\phi_i$ as the answer.

We now verify the correctness of the above query procedure.
First of all, we claim that $S \cap H = S_i \cup L$.
Indeed, we have $L \subseteq S \cap H$ by definition and $S_i \subseteq S \cap H$ because $a^*$ is below $\Xi_i$ (and hence below $h^*$) for all $a \in S_i$; this implies $S_i \cup L \subseteq S \cap H$.
To see $S \cap H \subseteq S_i \cup L$, let $a \in S \cap H$ be a point.
If $a^*$ is below $\Xi_i$, then $a \in S_i$.
Otherwise, there exists $v \in V$ such that $a^*$ is above $v$.
It follows that $a \in S \cap (H \cap H_v) \subseteq L$.
Therefore, $S \cap H \subseteq S_i \cup L$ and $S \cap H = S_i \cup L$.
With this observation in hand, we first show that the returned answer is a pair in $S \cap H$.
It suffices to show that both $\phi_i$ and $\psi$ are pairs in $S \cap H$.
The two points of $\phi_i$ are both in $S_i$ and hence in $S \cap H$.
To see $\psi$ is a pair in $S \cap H$, suppose $\psi = \psi_a$ for $a \in L$.
By definition, $\psi_a$ consists of $a$ and the nearest neighbor of $a$ in $P_a \backslash \{a\}$.
We have $a \in L \subseteq S \cap H$ and $P_a \subseteq L \subseteq S \cap H$, hence $\psi$ is a pair in $S \cap H$.
Next, we show that the returned answer is the closest pair in $S \cap H$.
Let $\phi^* = (a,b)$ be the closest-pair in $S \cap H$.
It suffices to show that $|\phi_i| \leq |\phi^*|$ or $|\psi| \leq |\phi^*|$.
If $a,b \in S_i$, then $|\phi_i| \leq |\phi^*|$ and we are done.
Otherwise, assume $a \notin S_i$ and thus $a \in L$, without loss of generality.
Since $\phi^*$ is the closest pair in $S \cap H$, we have $|\phi^*| \leq |\phi_i|$, which implies that the distance between $a$ and $b$ is at most $\delta$.
Therefore, $b \in P_a$.
Now we have $|\psi| \leq |\psi_a| \leq |\phi^*|$, which completes the proof of the correctness.
\smallskip

\noindent
\textbf{Analysis.}
We analyze the performance (space, query time, and preprocessing time) of our halfspace RCP data structure.
The space of the data structure is $O(n \log^{O(1)} n+R)$, because $\mathcal{D}(S)$ occupies $O(n)$ space, $\mathcal{D}'(S)$ occupies $O(n \log^{O(1)} n)$ space, and storing $\Xi_1,\dots,\Xi_R$ (with the associated pairs $\phi_1,\dots,\phi_R$) requires $O(R)$ space.
Next, we analyze the query time.
Determining the cell $\Xi_i$ takes $O(\log r)$ time by Lemma~\ref{lem-cutting}.
For each $v \in V$, reporting the points in $S \cap (H \cap H_v)$ takes $O(n^{1-1/d} \log^{O(1)}n + k_v)$ time where $k_v = |S \cap (H \cap H_v)|$.
We claim that $a^*$ intersects $\Xi_i$ for any $a \in S \cap (H \cap H_v)$.
Indeed, $a^*$ is below $h$ because $a \in H$ and is above $v$ because $a \in H_v$.
Thus, $a^*$ intersects the segment connecting $h^*$ and $v$.
Since $h^*,v \in \Xi_i$, $a^*$ intersects $\Xi_i$.
It follows that $k_v = O(n/r)$ by Lemma~\ref{lem-cutting}.
Furthermore, because $V = O(1)$, $L$ can be computed in $O(n^{1-1/d} \log^{O(1)}n + \sum_{v \in V}k_v) = O(n^{1-1/d} \log^{O(1)}n + n/r)$ time and $|L| = O(\sum_{v \in V}k_v) = O(n/r)$.
Once $L$ is obtained, $\phi'$ can be computed in $O((n/r) \log (n/r))$ time using the standard closest-pair algorithm.
For a point $a \in L$, reporting the points in $P_a$ takes $O(\log^{O(1)} n + m_a^{1-1/\lfloor d/2 \rfloor} \log^{O(1)} m_a + |P_a|)$ time where $m_a = |S \cap \Box_a|$, by Lemma~\ref{lem-BHRR}(b).
By exactly the same argument in the proof of Lemma~\ref{lem-number}, we have the following observation:
\begin{lemma}
$\sum_{a \in L} m_a = O(n)$ and $|P_a| = O(1)$ for all $a \in L$.
\end{lemma}

\newcommand{\PROOFOMIT}{
\textit{Proof.}
We first prove $\sum_{a \in L} m_a = O(n)$.
Consider a point $p \in S$.
Let $\Box_p$ be the hypercube centered at $p$ with side-length $2\delta$.
Note that $p \in P_a$ only if $a \in \Box_p$ for all $a \in L$.
Since $\phi'$ is the closest pair in $L$ and $|\phi'| \geq \delta$, we have $L \cap \Box_p = O(1)$ by the Pigeonhole Principle.
Therefore, only a constant number of points in $L$ is contained in $p$.
In other words, any point $p \in S$ is contained in $P_a$ for only a constant number of $a \in L$, which implies $\sum_{a \in L} m_a = O(n)$.
Next, we prove that $|P_a| = O(1)$ for all $a \in L$.
Since $S \cap H = S_i \cup L$ and $P_a = S \cap (\Box_a \cap H)$, it suffices to show $S_i \cap \Box_a = O(1)$ and $L \cap \Box_a = O(1)$.
Both facts follow from the Pigeonhole Principle readily.
Indeed, we have $S_i \cap \Box_a = O(1)$ because $\phi_i$ is the closest pair in $S_i$ and $|\phi_i| \geq \delta$.
We have $L \cap \Box_a = O(1)$ because $\phi'$ is the closest pair in $L$ and $|\phi'| \geq \delta$.
This completes the proof of $|P_a| = O(1)$.
\hfill $\Box$
\smallskip
}

\noindent
By the above lemma and H\"older's inequality, we have
\begin{equation*}
    \sum_{a \in L} m_a^{1-1/\lfloor d/2 \rfloor} \leq 
    O(n^{1-1/\lfloor d/2 \rfloor} |L|^{1/\lfloor d/2 \rfloor}) = O\left(\frac{n}{r^{1/\lfloor d/2 \rfloor}}\right),
\end{equation*}
which implies that computing all the $P_a$'s takes $O(n \log^{O(1)} n/r^{1/\lfloor d/2 \rfloor})$ time.
The pair $\psi$ can be directly obtained after knowing all the $P_a$'s.
Hence, the total query time is $O(\log r + n \log^{O(1)} n/r^{1/\lfloor d/2 \rfloor})$.
Finally, we analyze the preprocessing time.
The data structures $\mathcal{D}(S)$ and $\mathcal{D}'(S)$ can both be constructed in $O(n \log^{O(1)} n)$ time by Lemma~\ref{lem-SRR}(b) and~\ref{lem-BHRR}(b).
The cells $\Xi_1,\dots,\Xi_R$ can be computed in $O(nr^{d-1})$ time by Lemma~\ref{lem-cutting}.
So it suffices to show how to compute the pairs $\phi_1,\dots,\phi_R$ efficiently.
To this end, we build a simplex RCP data structure on $S$ as described in Theorem~\ref{thm-SRCP}, which takes $\tO(n^{(3d^2+1)/(2d^2+1)})$ time.
Fix $i \in \{1,\dots,R\}$ and let $V$ be the set of the $O(1)$ vertices of $\Xi_i$.
For $v \in V$, let $H_v'$ be the halfspace below the hyperplane $v^*$ in the primal space.
We claim that $S_i = S \cap (\bigcap_{v \in V} H_v')$.
To see this, consider a point $a \in S$.
We have $a \in S_i$ iff $a^*$ is below $\Xi_i$ iff $v$ is below $a^*$ for all $v \in V$, or equivalently, $a \in H_v'$ for all $v \in V$.
Thus, $S_i = S \cap (\bigcap_{v \in V} H_v')$.
We can then compute the closest pair $\phi_i$ in $S_i$ using the simplex RCP data structure with the query range $\bigcap_{v \in V} H_v'$ (as mentioned at the end of Section~\ref{sec-simplex}, our simplex RCP data structure can handle queries which are intersections of constant number of halfspaces).
Computing $\phi_i$ takes $O(n^{1-1/(2d^2)} \log^{O(1)} n)$ time, and hence computing all pairs $\phi_1,\dots,\phi_R$ takes $O(Rn^{1-1/(2d^2)} \log^{O(1)} n)$ time.
In sum, the preprocessing time of our halfspace RCP data structure is $O((nr^{d-1}+n^{(3d^2+1)/(2d^2+1)}+Rn^{1-1/(2d^2)})\log^{O(1)}n)$.
Setting $r = n^{1/d}$ gives: 
\begin{theorem} \label{thm-HRCP}
Given a set $S$ of $n$ points in $\mathbb{R}^d$, one can construct in $\tO(n^{2-1/(2d^2)})$ time a halfspace RCP data structure on $S$ with $\tO(n)$ space and $\tO(n^{1-1/(d\lfloor d/2 \rfloor)})$ query time.
\end{theorem}

\bibliography{my_bib.bib}

\appendix

\section{Ball RCP queries}

The same approach as in Theorem~\ref{thm-HRCP} also works for ball RCP queries, by applying the standard lifting transformation to map $d$-dimensional balls to $(d+1)$-dimensional halfspaces.
Note that, in general, ball RCP queries in $\mathbb{R}^d$ cannot be reduced to halfspace RCP queries in $\mathbb{R}^{d+1}$ using the lifting argument, as the lifting map does not preserve pairwise distances of the points.
However, our approach for handling halfspace queries (together with the lifting argument) can actually be applied to answer ball queries.
An easy way to see this is the following.
The lifting map $\rho:\mathbb{R}^d \rightarrow \mathbb{R}^{d+1}$ is defined as $(x_1,\dots,x_d) \mapsto (x_1,\dots,x_d,\sum_{i=1}^d x_i^2)$.
Let $\pi:\mathbb{R}^{d+1} \rightarrow \mathbb{R}^d$ be the projection map $(x_1,\dots,x_d,x_{d+1}) \mapsto (x_1,\dots,x_d)$.
Then $\pi \circ \rho = \text{id}_{R^d}$.
Define a distance function $f: \mathbb{R}^{d+1} \times \mathbb{R}^{d+1} \rightarrow \mathbb{R}^{d+1}$ as $f(a,b) = \lVert \pi(a)-\pi(b) \rVert_2$.
Clearly, we have $\lVert a-b \rVert_2 = f(\rho(a),\rho(b))$ for all $a,b \in \mathbb{R}^d$.
Therefore, the lifting argument reduces ball RCP search in $\mathbb{R}^d$ to halfspace RCP search in $\mathbb{R}^{d+1}$ under the distance function $f$.
Now observe that our halfspace RCP data structure works even under the distance function $f$.
Indeed, all of our RCP data structures in this paper only requires the distance function to satisfy two conditions: \textbf{(1)} closest-pair algorithms with near-linear time exist and \textbf{(2)} packing argument works.
It is clear that the distance function $f$ satisfies both of the requirements.
As such, we obtain a halfspace RCP data structure under the distance function $f$ with the same performance as in Theorem~\ref{thm-HRCP}, which in turn gives us a ball RCP data structure.

\begin{theorem} \label{thm-BRCP}
Given a set $S$ of $n$ points in $\mathbb{R}^d$, one can construct in $\tO(n^{2-1/(2(d+1)^2)})$ time a ball RCP data structure on $S$ with $\tO(n)$ space and $\tO(n^{1-1/((d+1)\lceil d/2 \rceil)})$ query time.
\end{theorem}





\end{document}